\documentclass[reprint,
onecolumn,
superscriptaddress,
nofootinbib,
aps,
pre,
]{revtex4-2}

\usepackage{array}
\usepackage{dcolumn}
\usepackage{amsmath,amssymb,graphicx,comment,bm,mathtools}
\usepackage[normalem]{ulem}
\usepackage[dvipsnames]{xcolor}
\usepackage{url}
\usepackage{tikz}
\usepackage{hyperref}

\begin{document}

\title{
An effective semilocal model for wave turbulence in 2D nonlinear optics
}

\author{Jonathan Skipp}
\email{j.skipp@aston.ac.uk}

\affiliation{Department of Mathematics, College of Engineering and Physical Sciences, Aston University, Aston Triangle, Birmingham, B4 7ET, UK}

\author{Jason Laurie}
\email{j.laurie@aston.ac.uk}

\affiliation{Department of Mathematics, College of Engineering and Physical Sciences, Aston University, Aston Triangle, Birmingham, B4 7ET, UK}

\author{Sergey Nazarenko}
\email{Sergey.nazarenko@univ-cotedazur.fr}

\affiliation{Universit\'e C\^ote d'Azur, CNRS-Institut de Physique de Nice, 17 Rue Julien Laupr\^etre, 06200, Nice, France}

\begin{abstract}
The statistical evolution of ensembles of random, weakly-interacting waves is governed by wave kinetic equations. To simplify the analysis, one frequently works with reduced differential models of the wave kinetics. However, the conditions for deriving such reduced models are seldom justified self-consistently. Here, we derive a reduced model for the wave kinetics of the Schr{\"o}dinger-Helmholtz equations in two spatial dimensions, which constitute a model for the dynamics of light in a spatially-nonlocal, nonlinear optical medium. This model has the property of sharply localising the frequencies of the interacting waves into two pairs, allowing for a rigorous and self-consistent derivation of what we term the semilocal approximation model (SLAM) of the wave kinetic equation. Using the SLAM, we study the stationary spectra of Schr{\"o}dinger-Helmholtz wave turbulence, and characterise the spectra that carry energy downscale, and waveaction upscale, in a forced-dissipated setup. The latter involves a nonlocal transfer of waveaction, in which waves at the forcing scale mediate the interactions of waves at every larger scale. This is in contrast to the energy cascade, which involves local scale-by-scale interactions, familiar from other wave turbulent systems and from classical hydrodynamical turbulence. 
\end{abstract}

\maketitle

\section{Introduction}\label{sec:intro}

Wave turbulence is the statistical theory of large ensembles of random, weakly nonlinear, dispersive waves \cite{zakharov1992book, nazarenko2011book}. 
Accordingly, when developing the wave turbulence description of a physical system, one is most frequently concerned with the mean square of the wave intensity: the waveaction spectrum. 
The equation of motion for the spectrum is known as the wave kinetic equation (WKE), and describes the irreversible evolution of the spectrum  over nonlinear timescales (which are long compared to the linear wave period), due to resonant $M$-wave interactions. 
	For a system in $d$ spatial dimensions, the WKE involves an integration over $\mathbb{R}^{(M-1) d}$, constrained to the resonant manifold of interacting waves.
	The complexity of this so-called collision integral makes solving the WKE a challenging task in general. 
	Nonetheless, certain analytic techniques exist, in particular the Zakharov-Kraichnan transform that allows one to find the Kolmogorov-Zakharov (KZ) cascade	spectra \cite{zakharov1965weak}. These are stationary solutions of the WKE on which, for many systems, the dynamical invariants cascade with constant flux through spatial scales, via a self-similar, spectrally local (scale-by-scale) transfer, analogous to the Kolmogorov energy spectrum in classical hydrodynamics. 
	The stationary spectrum of thermodynamic equilibrium---the Rayleigh-Jeans (RJ) spectrum---can also be derived trivially as the spectrum on which the collision integral has an integrand that vanishes pointwise.

	In a seminal paper, Dyachenko et.\ al.\ \cite{dyachenko1992optical} demonstrated that the collision integral can be greatly simplified if one makes the \emph{ad-hoc} assumption that the wave interaction coefficient is sharply peaked, so that all $M$ waves taking part in interactions have approximately the same frequency. 
	This assumption, which we will refer to as superlocality, allows one to reduce the collision integral to a differential operator.
	The resulting equation---the differential approximation model (DAM)---preserves a great deal of the structure of the original WKE, namely its conserved quantities, the degree of nonlinearity with respect to the spectrum, its scaling with frequency, and, as a result of the latter, the stationary RJ and KZ solutions. 
	DAMs are the wave turbulence equivalent of the Leith model of classical hydrodynamics \cite{leith1968diffusion, connaughton2004warm}. 
	
	Being differential equations, DAMs are much easier to work with than the collision integrals from which they are derived. 
	They have been used in a wide variety of physical systems to examine topics such as the stationary RJ and KZ wave turbulence spectra \cite{dyachenko1992optical, zakharov1999diffusion, galtier2000WT_MHD, galtier2010diffusioneqs_MHD, proment2012warm, Laurie2012_1DOpticalWT, skipp2020wave}, thermalisation at the end of a cascade spectrum \cite{boffetta2009modeling}, the crossover from strong to weak turbulence \cite{lvov2007bottleneck}, and the nature of transient spectra before the KZ spectra are established, including the anomalous scaling of spectral fronts \cite{connaughton2003nonstationary, thalabard2015anomalous, galtier2019nonlinear, grebenev2020steady}.

	The reduction of a WKE to a DAM is predicated on the assumption of superlocality. 
	However, this assumption is rarely justified in the cases where DAMs are applied, indeed Dyachenko et.\ al.\ \cite{dyachenko1992optical} introduced the DAM in the context of the cubic nonlinear Schr\"odinger equation, whose interaction coefficient is a constant across Fourier space. Furthermore, DAMs are often constructed heuristically, based on the scaling properties of the interaction coefficient, with the desired stationary solutions and degree of nonlinearity built in. 
	To our knowledge, there has been no rigorous derivation of the DAM for any system whose interaction coefficient has the required properties to justify any locality assumption.

In this paper, we derive such a reduced model for the wave kinetics of the Schr{\"o}dinger-Helmholtz equations (SHE). 
	We introduce the SHE in Sec.\ \ref{sec:SHE}, along with their physical context, the dynamical invariants that they conserve (namely energy and waveaction), their WKE, and the directions in Fourier space that their invariants flow during the wave kinetic evolution.
	The SHE are of interest to us because they comprise the first system studied in the wave turbulence context in which the spectral locality of interactions arises naturally from the functional form of the interaction coefficient. In fact, the locality manifested by the SHE is one in which distinct pairs of interacting waves are localised in frequency space. 
	We refer to the latter as a semilocal, as opposed to a superlocal, limit.	
	We exploit this property in Sec.\ \ref{sec:SLAM_derivation} to reduce the kinetic equation of the SHE to a simpler model, in the same spirit as the derivation of the DAM in Ref.\ \cite{dyachenko1992optical}. 
	It transpires that the semilocality property allows the collision integral to be reduced to an integro-differential operator, rather than a purely differential one. The resulting reduction of the WKE we term the semilocal approximation model (SLAM).

	Analysis of the SLAM allows us to extract the stationary spectra of the WKE, including prospective candidates for	the KZ cascade spectra, in Sec.\ \ref{sec:stationarysols}. 
However, we demonstrate that the KZ waveaction cascade spectrum is pathological, as it leads to a divergence of the SLAM at high frequency.
	Furthermore, in Sec.\ \ref{sec:warm} we show how the KZ spectra lead to flux directions that are inconsistent with the more general argument we present in Sec.\ \ref{subsubsec:fjortoft}, requiring us to reconsider the spectra that establish the turbulent transport of dynamical invariants across scales of the system.
	We proceed to find the true waveaction flux spectrum in Sec.\ \ref{sec:inverse_cascade}, and conclude that both stationary solutions that describe the flux of energy on the one hand, and waveaction on the other, are very closely related to the RJ equilibrium spectrum. 	
	The waveaction flux spectrum, which carries waveaction to large scales, is dominated by nonlocal interactions, with waves at the forcing scale mediating wave interactions at all larger scales. By contrast, the energy flux spectrum, which carries energy to small scales, has local scale-by-scale interactions.
	We start by introducing the SHE in the next section.

\section{Schr\"odinger-Helmholtz equation}\label{sec:SHE}

The SHE consist of a nonlinear Schr\"odinger equation for the dynamical variable $\psi({\bf x},t)  \in \mathbb{C}$,
\begin{subequations}\label{eq:SHE}
\begin{align}\label{eq:SHE_schro}
i \frac{\partial \psi}{\partial t} + \nabla^2\psi - V(\psi)\psi &=0,
\end{align}
coupled, via the potential $V(\psi) \in \mathbb{R}$, to the Helmholtz equation,
\begin{align}\label{eq:SHE_Helm}
\nabla^2V-\Lambda V &= 
\alpha
|\psi|^2.
\end{align}
\end{subequations}
We take the spatial domain to be a $d$-dimensional periodic box of side length $L$.
The SHE \eqref{eq:SHE} are a spatially nonlocal\footnote{
											The spatial nonlocality originates from the 
											inversion of the Helmholtz operator in Eq.\ \eqref{eq:SHE_Helm}.
											The term ``local'' is used here in a different sense to the locality of 
											interactions in frequency space, to which the semilocal approximation refers.
	}
version of the familiar cubic nonlinear Schr\"odinger equation (NLSE, also known as the Gross-Pitaevskii equation),
\begin{align}
\label{eq:NLS}
i\frac{\partial \psi}{\partial t} + \nabla^2\psi \pm |\psi|^2 \psi =0.
\end{align}
The NLSE is obtained from Eqs.\ \eqref{eq:SHE} by sending $
\alpha,
\Lambda \to \infty$ in such a way  that $
\alpha
/\Lambda$ remains constant, and renormalising $\psi$.

The physical applications of the SHE were discussed in Ref.\ \cite{skipp2020wave}; in brief, for $d=3$ they describe so-called Fuzzy Dark Matter \cite{lee1996galactic, SuarezBECDM2014,  mocz2017galaxyI} in a universe with cosmological constant $\Lambda$.
In $d=2$, Eqs.\ \eqref{eq:SHE} describe the perpendicular dynamics of laser light in a  thermo-optic or elasto-optic nonlinear medium \cite{peccianti2003optical, khoo2007liquidcrystalsbook, paredes2020optics}. In the optics case, $\Lambda$ is the normalised Kerr coefficient of spatially-local interactions.
The dynamical variable $\psi({\bf x},t)$ represents, respectively, the wavefunction of the putative dark matter boson, or the envelope of the electric field.
Here we restrict ourselves to the $d=2$ case.

Also closely associated with the SHE are the Schr\"odinger-Newton equations (SNE) \cite{ruffini1969systems, diosi1984gravitation, penrose1996gravity},
\begin{subequations}\label{eq:SNE}
\begin{align}
		i\frac{\partial \psi}{\partial t} + \nabla^2\psi - V(\psi)\psi =0,  \label{eq:SNE_schro} \\
		\nabla^2V = 
		\alpha
		|\psi|^2  \label{eq:SNE_Pois},
\end{align}
\end{subequations}
which are obtained by formally setting $\Lambda = 0$ in the SHE. However, as discussed in Ref.\ \cite{skipp2020wave}, the SNE are ill-posed in periodic settings, or when one wants to describe fluctuations over an infinite, static background, because that background does not solve the Poisson equation \eqref{eq:SNE_Pois}, c.f.\ the ``Jeans swindle''  \cite{BinneyTremaine1987_GalacDynBook}. Non-trivial dynamics are recovered by introducing a spatially local term to the left-hand side of Eq.\ \eqref{eq:SNE_Pois}, i.e.\ moving to the SHE  \cite{kiessling2003jeans}.
This fact is reflected in the divergence of the SLAM when we set $\Lambda=0$, see Sec.\ \ref{subsec:cf2020}.

\subsection{Hamiltonian and invariants of the SHE}
\label{subsec:inv_SHE}

Most commonly in wave turbulence, the equation of motion under study can be derived via Hamilton's equation $i\partial_t \psi = \delta H / \delta \psi^*$. Equations \eqref{eq:SHE} are no exception, with the Hamiltonian functional being
\begin{align}
\label{eq:SHE_Ham}
H = \int   |\nabla \psi|^2  \,  d{\bf x} 
			+ \int \frac{
				\alpha
				}{2} \left[ (\nabla^2 - \Lambda)^{-1/2} |\psi|^2 \right]^2 \, d{\bf x}.
\end{align}
The first term on the right-hand side of Eq.\ \eqref{eq:SHE_Ham} is the quadratic energy, associated with the free propagation of dispersive waves. The second term is the energy contribution due to the nonlinear interaction of waves via the spatially-nonlocal potential $V(\psi) = 
\alpha
(\nabla^2 - \Lambda)^{-1}|\psi|^2$ that solves Eq.\ \eqref{eq:SHE_Helm}. The operator $(\nabla^2-\Lambda)^{-q}$ with $q$ rational and positive, is to be understood as a formal power series, and is made concrete in its Fourier-space representation (the latter is found in Ref.\ \cite{skipp2020wave}).

In the absence of forcing and dissipation (see below),
the Hamiltonian $H$ is conserved under the evolution via the SHE, and is strictly positive. The other positive invariant is the waveaction (a.k.a. number of particles, in reference to the application to Bosonic systems),
\begin{align}
\label{eq:SHE_N}
N = \int   |\psi|^2   \,  d{\bf x}.
\end{align}
The momentum ${\bf P} = (i/2) \int  (\psi \nabla\psi^*  -  \psi^* \nabla\psi)  \, d{\bf x}$ is yet another conserved quantity. However, not being sign-definite, it plays no role in the argument regarding the invariant cascade directions, see Sec.\ \ref{subsubsec:fjortoft}. The momentum will not feature in the work we carry out in this paper.

\subsection{SHE in Fourier space}
\label{subsec:SHE_Fourier}

In Fourier space Eqs.\ \eqref{eq:SHE} become an equation of motion for $\psi_{\bf k}(t)=(1/L)^d \int \psi({\bf x},t) \exp(-i{\bf k}\cdot{\bf x}) d{\bf x}$, the Fourier series coefficient of $\psi({\bf x},t)$ for the wave mode with wavevector ${\bf k}$:
\begin{equation}
	\label{eq:SHE_Fourier}
	i \frac{\partial\psi_{\bf k}}{\partial t} 
	- k^2 \psi_{\bf k} 
	- \!\!\! \sum_{{\bf k}_1 , {\bf k}_2 ,  {\bf k}_3} \!\!\!
			W^{12}_{3{\bf k}} \psi_{{\bf k}_1} \psi_{{\bf k}_2} \psi_{{\bf k}_3}^* \delta^{12}_{3{\bf k}}
	=0.
\end{equation}
The quantity 
\begin{align}\label{eq:W123k}
	W^{12}_{3{\bf k}} = - \frac{
		\alpha
	}{2}    \left[ \frac{1}{|{\bf k}_1-{\bf k}|^2+\Lambda}   +    \frac{1}{|{\bf k}_2-{\bf k}|^2+\Lambda} \right]
	\end{align} 
is the interaction coefficient for the SHE. It gives the strength of nonlinear interactions between tetrads of waves that satisfy ${\bf k}_1+{\bf k}_2-{\bf k}_3-{\bf k}=0$, 
to which the sum in Eq.\ \eqref{eq:SHE_Fourier} is constrained via the Kronecker delta 
$\delta^{12}_{3{\bf k}} \coloneqq \delta^{{\bf k}_1,{\bf k}_2}_{{\bf k}_3,{\bf k}}$. 
The functional dependence of $W^{12}_{3{\bf k}}$ on the wavevectors is indicated in the super- and subscript indices, and its functional form is obtained by writing Eqs.\ \eqref{eq:SHE_schro} and \eqref{eq:SHE_Helm} in Fourier space and combining. 
Using the Kronecker delta $\delta^{12}_{3{\bf k}}$, we establish the symmetries $W^{12}_{3{\bf k}}  =  W^{21}_{3{\bf k}}  =  W^{12}_{{\bf k}3}  =  \left( W_{12}^{3{\bf k}} \right)^*$, which are required by the fact that the Hamiltonian \eqref{eq:SHE_Ham} is real.

\subsubsection{Freely-evolving vs.\ forced-dissipated systems}

In a closed system, namely one with no sources or sinks of dynamical invariants, Eq.\ \eqref{eq:SHE_Fourier}  (equivalently Eq.\ \eqref{eq:SHE}) describes the conservative dynamics of the field $\psi_{\bf k}(t)$ (equivalently $\psi({\bf x},t)$) evolving freely from an initial condition. Alternatively, one can consider a system in which invariants are injected into the system by some forcing mechanism, and removed from it by dissipation. These mechanisms manifest as extra terms $F_{\bf k}$ and $D_{\bf k}$, respectively, on the right-hand side of Eq.\ \eqref{eq:SHE_Fourier} (and their Fourier inverses on the right-hand side of Eq.\ \eqref{eq:SHE_schro}). Analogous terms appear on the right-hand side of the WKE \eqref{eq:kinetic} in the forced-dissipated case.

In turbulence theory, one usually considers forcing to be isotropic (depending only on $k\coloneqq|{\bf k}|$), statistically time-independent, and to act in a narrow band of lengthscales around a characteristic wavenumber $k_f$. $F_{\bf k}$ is then a stochastic term supported within the forcing range and negligible elsewhere. Dissipation is usually significant at the largest and/or smallest lengthscales of the system, and negligible elsewhere. In this paper, we assume both large-scale and small-scale isotropic dissipation, with respective characteristic wavenumbers $k_{d-}$ and $k_{d+}$. 

Furthermore, we take the forcing and dissipation scales to be widely separated: $k_{d-} \ll k_f \ll k_{d+}$. Lengthscales in between, but far from, the forcing and dissipation scales are known as the inertial ranges. Inside the inertial ranges the dynamics are conservative, described by Eq.\ \eqref{eq:SHE_Fourier} (equivalently \eqref{eq:SHE_schro}, and when we consider wave turbulence, the WKE \eqref{eq:kinetic}) with zero right-hand side. The role of forcing and dissipation is to provide a source and sink of invariants on either side of the inertial ranges, setting up a flux of invariants through them. 
We discuss the directions of these fluxes in section \ref{subsubsec:fjortoft}.

\subsection{Wave kinetic equation}
\label{subsec:WKE}

Wave turbulence is primararily concerned with the waveaction spectrum $n_{\bf k}(t) = (L/2\pi)^d \langle |\psi_{\bf k}(t)|^2 \rangle$.
The operator $\langle \cdot \rangle$ denotes an average over a statistical ensemble of realisations, starting from independent initial conditions $\psi_{\bf k}(0)$, with phases uniformly distributed in $[0,2\pi)$, and identically distributed amplitudes.
Taking the domain size $L\to\infty$, and then assuming weak nonlinearity, one can derive the following WKE describing the evolution of the spectrum at intermediate times due to the nonlinear 4-wave interactions of the $2 \leftrightarrow 2$ type \cite{nazarenko2011book}:
\begin{align}\label{eq:kinetic}
\frac{\partial n_{\bf k} }{\partial t} = 4\pi \int     |W^{12}_{3{\bf k}}|^2    \delta^{12}_{3{\bf k}}    \delta(\omega^{12}_{3{\bf k}})   
								n_{1}  n_{2}  n_{3}  n_{\bf k} 
								\left[  \frac{1}{n_{\bf k}} + \frac{1}{n_{3}}-\frac{1}{n_{1}}-\frac{1}{n_{2}}  \right]  
								d{\bf k}_1\,  d{\bf k}_2\, d{\bf k}_3.
\end{align}
The right-hand side of Eq.\ \eqref{eq:kinetic} is the collision integral, and is taken across the joint ${\bf k}$-space $\mathbb{R}^2 \times \mathbb{R}^2 \times \mathbb{R}^2$.
Here, $\delta^{12}_{3{\bf k}} \coloneqq \delta({\bf k}_1 + {\bf k}_2 - {\bf k}_3 - {\bf k})$ and $\delta(\omega^{12}_{3{\bf k}})$ are 
Dirac delta functions that constrain interacting wave tetrads to the resonant manifold defined by
\begin{subequations} \label{eq:resonance}
\begin{align}
{\bf k}_1   +   {\bf k}_2   -   {\bf k}_3   -   {\bf k} = 0,   \label{eq:k_resonance}\\
\omega^{12}_{3{\bf k}} \coloneqq \omega_1 + \omega_2 - \omega_3 - \omega_{\bf k} = 0. \label{eq:om_resonance}
\end{align}
\end{subequations}
Here $\omega_{\bf k} = k^2$ is the linear dispersion relation. 
We have also used the shorthand notation $n_i = n_{{\bf k}_i}, \omega_i = \omega_{{\bf k}_i}$ etc.\ for $i=1,2,3$.


Inspecting Eq.\ \eqref{eq:W123k}, we see manifestly that the interaction coefficient $W^{12}_{3{\bf k}}$ decays rapidly when all wavevectors are very different. 
The first term in $W^{12}_{3{\bf k}}$ becomes dominant when ${\bf k}_1 \to {\bf k}$.
By Eq.\ \eqref{eq:k_resonance} we then have ${\bf k}_2 \to {\bf k}_3$. If we also have $\Lambda \ll k^2$, then the interaction coefficient becomes sharply peaked in the joint ${\bf k}$-space where ${\bf k}_1 \approx {\bf k}$ and ${\bf k}_3 \approx {\bf k}_2$, with the latter following from the above symmetries.
	 Likewise, if the second term in $W^{12}_{3{\bf k}}$ is dominant then it becomes peaked over ${\bf k}_1 \approx {\bf k}_3$ and ${\bf k}_2 \approx {\bf k}$.
	 These pairings are equivalent to the first pairings by exchange of dummy variables, as $W^{12}_{3{\bf k}}$ always appears under a sum or integral. 

	Thus, the 4-wave interactions responsible for evolution of the system under the SHE \eqref{eq:SHE}, and therefore the corresponding WKE \eqref{eq:kinetic}, are dominated by interactions in which ${\bf k}_1 \approx {\bf k}$ and ${\bf k}_3 \approx 
	{\bf k}_2
	$. 
	This property of the interaction coefficient, of picking out dominant interactions when pairs of wavevectors become equal, we refer to as semilocality.
	It is this property that will allow the collision integral to be reduced to a simpler operator. 
We will retain the possibility that ${\bf k}_1 \not\approx {\bf k}_2$, so the reduction will be to an integro-differential, rather than a purely differential, operator. 

At this point we note that taking the NLSE limit $
\alpha, \Lambda \to\infty$ with $
\alpha
/\Lambda\to\text{const.}$ sends $W^{12}_{3{\bf k}} \to \pm 1$. In this case there is no natural pairing of $({\bf k}_1 , {\bf k})$ and $({\bf k}_2 , {\bf k})$, and we lose the semilocality property. We return to this point in Sec.\ \ref{subsec:cf2020}.

\subsubsection{Invariants of the wave kinetic equation}
\label{subsubsec:invariants}

In general, WKEs of the $M/2 \leftrightarrow M/2$ type ($M$ being an even integer denoting the order of the resonant wave interaction), such as Eq.\ \eqref{eq:kinetic}, conserve the two quadratic invariants
\begin{align}
\label{eq:invariants_int_k}
E = \int \omega_{\bf k} n_{\bf k} \, d{\bf k}
\quad \text{and} \quad
N = \int n_{\bf k} \, d{\bf k}.
\end{align}
Comparing with Eq.\ \eqref{eq:SHE_N} we see that $N$ is the total waveaction, expressed in Eq.\ \eqref{eq:invariants_int_k} as an integral over Fourier space. Both the original SHE and the WKE derived from them conserve the waveaction exactly.
By contrast, $E$ is the Fourier-space representation of the quadratic energy (first term on the right-hand side of Eq.\ \eqref{eq:SHE_Ham}). Recall that the WKE is derived under the assumption of weak nonlinearity. Under this condition, $E$ will be the leading contribution to the total Hamiltonian $H$, i.e.\ the original equations of motion conserve $H$ exactly, while their WKE conserves $H$ asymptotically, and $E$ exactly. Here, we are interested in the quantities that are conserved by the WKE and its reduced model, the SLAM. Therefore, we will simply refer to $E$ as the energy hereafter.

The interaction coefficient $W^{12}_{3{\bf k}}$ is unchanged under global rotations. 
We further assume that when the system is forced and dissipated, it is done so in a spatially homogeneous and isotropic manner.
Therefore, we expect that the spectrum $n_{\bf k}$ will be isotropic, depending only on $|{\bf k}|$, or equivalently on frequency. Accordingly, we can consider the spectrum as a function of either ${\bf k}$, or frequency $\omega$ at that value of ${\bf k}$, via the dispersion relation $\omega=k^2$. Namely, we adopt the notation $n_{\omega_i} \coloneqq n({\bf k}_i(\omega_i) )= n_{{\bf k}_i} \eqqcolon  n_i$. 
Converting the ${\bf k}$-space integrals in Eqs.\ \eqref{eq:invariants_int_k} into integrals over $\omega$, the invariants of the WKE become, for a 2D isotropic spectrum,
\begin{align}
\label{eq:invariants}
E = \pi \int \omega n_\omega \, d\omega
\quad \text{and} \quad
N = \pi \int n_\omega \, d\omega.
\end{align}

\subsubsection{Flux directions - the Fj{\o}rtoft argument}
\label{subsubsec:fjortoft}

The action of the WKE is to redistribute the spectral density of $E$ and $N$ across ${\bf k}$-space.
The qualitative manner of this redistribution
is predicted by the argument of Fj{\o}rtoft \cite{fjortoft1953changes}. 
This argument is recapitulated in many places in the wave turbulence literature, see for example Refs.\ \cite{Laurie2012_1DOpticalWT, skipp2020wave} for its application to the forced-dissipated SHE, and Ref.\ \cite{nazarenko2011book} for a version of the argument in freely-evolving systems.

We restrict our discussion to the isotropic case, which allows us to elide from ${\bf k}$-space to $\omega$-space via the dispersion relation, and speak of scales when referring to frequencies.
The conclusion of the 
Fj{\o}rtoft
argument is that the presence of each invariant constrains how the 
$\omega$-space
distribution of the other invariant can evolve, so that the bulk of each invariant moves to the sector of 
$\omega$-space
where its spectral density dominates.
For the SHE this means that the majority of the energy $E$, which has a spectral density of $\omega = k^2$, moves towards high 
$\omega$, 
whereas most of the waveaction $N$, having a spectral density of $1$, moves towards low
$\omega$.

More specifically, for a freely-evolving system the Fj{\o}rtoft argument predicts that the invariant densities are redistributed by the WKE so that the centroid of $E$, defined as 
$(\pi/E)\int\!\omega^2 n_\omega \, d\omega$,  
moves towards small scales, while the $N$-centroid,
$(\pi/N)\int\!\omega n_\omega \, d\omega$,
moves towards the largest scale in the system \cite{nazarenko2011book}. The total $E$ and $N$ of course remain constant during the evolution.

In a forced-dissipated system, $E$ and $N$ are injected at the intermediate forcing scale $\omega_f$, and transported by the WKE until they reach the dissipation scales at $\omega_{d-}$ (large scale), and $\omega_{d+}$ (small scale), where they are removed. We assume a wide scale separation $\omega_{d-} \ll \omega_f \ll \omega_{d+}$.
We further assume that the system has reached a non-equilibrium stationary state in which forcing and dissipation are continuous, and the rate of dissipation has adjusted to match the rate of forcing, so that the total $E$ and $N$ remain constant. We conjecture that this steady state condition is universal, independent of the detailed manner of forcing and dissipation, and will be attained from a wide range of initial conditions after a transient phase.
In these circumstances, the Fj{\o}rtoft argument predicts that most ofthe energy injected at $\omega_f$ will 
be transferred with constant positive
energy flux $P$ through the direct inertial range 
(namely,
scales $\omega$ such that $\omega_f \ll \omega \ll \omega_{d+}$), to be dissipated at small scales around $\omega_{d+}$. Likewise, most of the waveaction injected at $\omega_f$ will  
be transferred with constant negative
waveaction flux $Q$ through the inverse inertial range ($\omega_{d-}\ll\omega\ll\omega_f$), until it is dissipated at large scales near $\omega_{d-}$. 
This scenario, of one invariant moving to small scales and the other moving to large scales, is common to all wave turbulence systems with two quadratic invariants, and also 2D hydrodynamic turbulence \cite{nazarenko2011book}. 
It is referred to in the literature as the dual cascade, although the term ``cascade'' usually implies a scale-by-scale flux of invariants in which all the waves taking part in the transfer are localised in $\omega$-space. 
We will see that for the SLAM, the direct transfer of energy involves entirely local interactions, whereas the inverse waveaction transfer is nonlocal, involving waves at $\omega_f$ participating in every tetrad of interacting waves throughout the inertial range.
In this paper we will still speak of the direct cascade of energy, and inverse cascade of waveaction, which together make up the dual cascade, having made this caveat about the (non)locality of interactions in these cascades.
	
The Fj{\o}rtoft argument is premised only on having positive-definite integral invariants, which are quadratic in wave amplitude, but which have different spectral densities, and on having widely-separated forcing and dissipation scales. 
Having such parsimonious assumptions, the predictions of the Fj{\o}rtoft argument are robust, and must be recovered by any subtler manipulation of the WKE. More concretely, once we derive the SLAM, we can look for its stationary solutions that realise the dual cascade, but these solutions must have fluxes $P>0$ and $Q<0$ in their respective inertial ranges, to correspond to the predictions of the Fj{\o}rtoft argument.
On the other hand, the argument makes no assumptions about the locality of interactions in $\omega$-space. This must be determined by examining individual candidate solutions, which is particularly straightforward in the SLAM.

\section{Derivation of the semilocal approximation model}
\label{sec:SLAM_derivation}

To derive the SLAM, we follow the initial strategy set out in Ref.\ \cite{dyachenko1992optical} for deriving the DAM. 
First, we multiply Eq.\ \eqref{eq:kinetic} by an arbitrary test function $\varphi_{\bf k} = \varphi({\bf k})$, integrate with respect to ${\bf k}$, and use the resulting symmetries of the integrand to split it into four pieces:
\begin{align}
\int \varphi_{\bf k}\frac{\partial n_{\bf k}}{\partial t} \, d{\bf k}
	&=	4\pi \int 	\varphi_{\bf k}    |W^{12}_{3{\bf k}}|^2    \delta^{12}_{3{\bf k}}    \delta(\omega^{12}_{3{\bf k}})
				n_{1}  n_{2}  n_{3}  n_{\bf k} 		
				\left[  \frac{1}{n_{\bf k}} + \frac{1}{n_{3}} - \frac{1}{n_{1}} - \frac{1}{n_{2}}  \right]
				d{\bf k}_1\,  d{\bf k}_2\, d{\bf k}_3 \, d{\bf k}      		  \nonumber \\
	&=	\pi \int		\left[  \varphi_{\bf k} + \varphi_{3} - \varphi_{1} - \varphi_{2}  \right]    
				|W^{12}_{3{\bf k}}|^2    \delta^{12}_{3{\bf k}}    \delta(\omega^{12}_{3{\bf k}})
				n_{1}  n_{2}  n_{3}  n_{\bf k}  
				\left[  \frac{1}{n_{\bf k}} + \frac{1}{n_{3}} - \frac{1}{n_{1}} - \frac{1}{n_{2}}  \right] 
				d{\bf k}_1\, d{\bf k}_2\, d{\bf k}_3\, d{\bf k}.      \label{eq:int_fndot}
\end{align}

Next we assume that the spectra $n_{\bf k}$ and test functions $\varphi_{\bf k}$ are isotropic, and consider both as functions of frequency (see Sec.\ \ref{subsubsec:invariants}).

At this point, following the discussion after Eq.\ \eqref{eq:om_resonance}, we make the semilocality assumption that ${\bf k}_1 \approx {\bf k}$, and hence ${\bf k}_3 \approx {\bf k}_2$, but retain the possibility of ${\bf k}_1$ and ${\bf k}_2$ being distinct. This is in contrast to the procedure of Dyachenko et al.\ \cite{dyachenko1992optical}, who assume that \emph{all} interactions are superlocal in frequency space.

\begin{figure}[!h]
	\begin{center}
	\includegraphics[width=0.75\columnwidth]{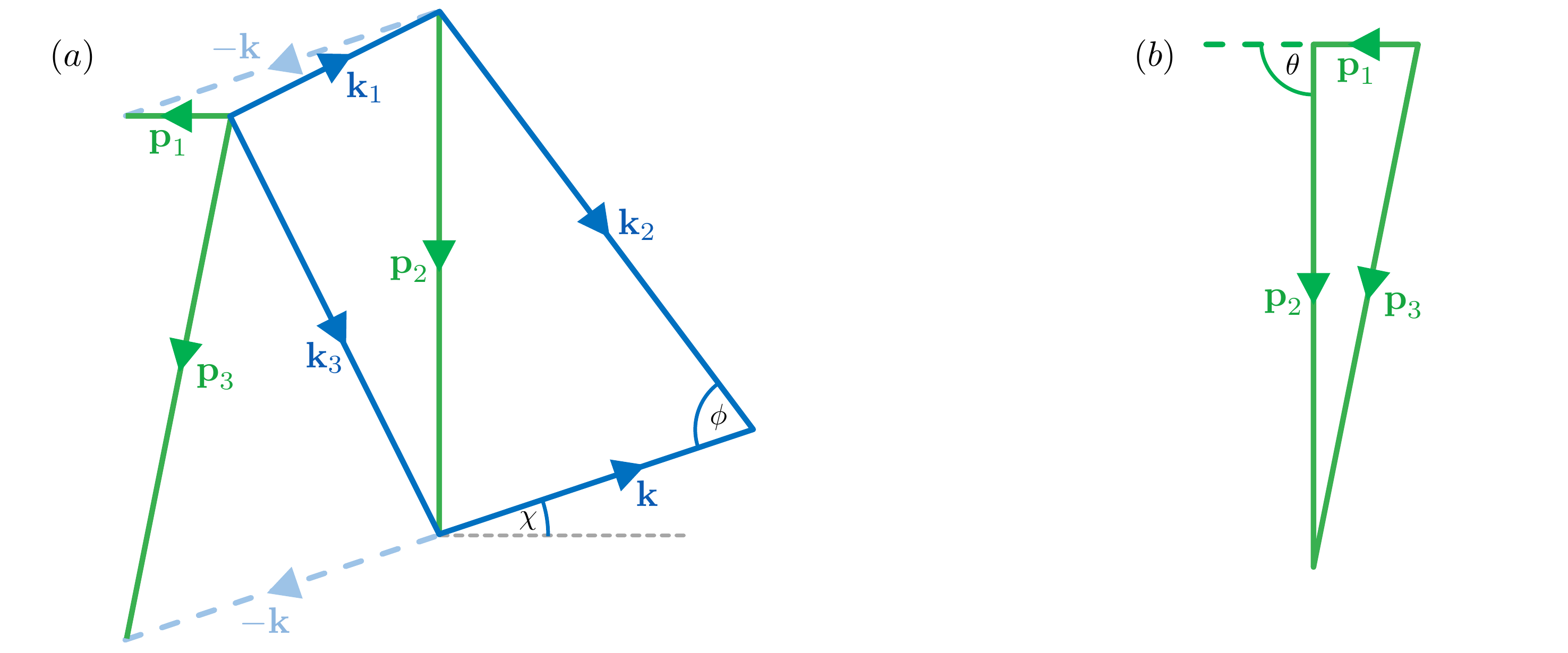}
	\caption{\label{fig:angles}
		Geometry of wavevectors and definition of angles involved in deriving the SLAM.
		(a) A tetrad of wavevectors $\{{\bf k}_1, {\bf k}_2, {\bf k}_3, {\bf k}\}$ that satisfy the resonant conditions Eqs.\ \eqref{eq:resonance} are shown in blue. Note the semilocal approximation ${\bf k}_1\approx {\bf k}$ and ${\bf k}_3\approx {\bf k}_2$. 
		The difference vectors $\{{\bf p}_i\coloneqq {\bf k}_i-{\bf k}\}$ are shown in green.
		(b) Triangle formed by $\{{\bf p}_i\}$. Note that due to the frequency resonance condition $\delta(\omega^{12}_{3{\bf k}})$ in Eq.\ \eqref{eq:kinetic}, only wavevectors with difference vectors that form a right-angle triangle, i.e. $\theta=\pi/2$, will contribute to the collision integral, see Appendix \ref{app:triangle}.
	}
	\end{center}
\end{figure}

Taylor expanding the terms in square brackets in Eq.\ \eqref{eq:int_fndot} up to first order in frequency, and using Eq.\ \eqref{eq:om_resonance}, we have
\begin{equation}
	\begin{aligned}
	\left[ \frac{1}{n_{\bf k}} + \frac{1}{n_{3}} - \frac{1}{n_{1}} - \frac{1}{n_{2}} \right]
		&\approx  \partial_\omega {n_\omega}^{-1}(\omega- \omega_1) - \partial_{\omega_2} n_{\omega_2}^{-1}(\omega_2- \omega_3)  \\
		&= 			\left(  \partial_\omega n_\omega^{-1}-\partial_{\omega_2} n_{\omega_2}^{-1}  \right)(\omega- \omega_1) \\
		&= 			\left(  \partial_\omega n_\omega^{-1}-\partial_{\omega_2} n_{\omega_2}^{-1}  \right)({\bf k}- {\bf k}_1)\cdot({\bf k}+ {\bf k}_1) \\
		&\approx  \left(  \partial_\omega n_\omega^{-1}-\partial_{\omega_2} n_{\omega_2}^{-1}  \right)( - {\bf p}_1)\cdot 2{\bf k},	
		\label{eq:Taylorexpns}
	\end{aligned}
\end{equation}
where the difference vectors ${\bf p}_i \coloneqq {\bf k}_i-{\bf k}$, shown in green in Fig.\ \ref{fig:angles}(a).
Similarly,
\begin{align*}
\left[ {\varphi_{\bf k}} + {\varphi_{3}} - {\varphi_{1}} - {\varphi_{2}} \right]  
		\approx  \left(  \partial_\omega \varphi_\omega-\partial_{\omega_2} \varphi_{\omega_2}  \right)( - {\bf p}_1)\cdot 2{\bf k}.
\end{align*}
Therefore, Eq.\ \eqref{eq:int_fndot} simplifies to
\begin{align*}
\int \varphi_{\bf k}\frac{\partial  n_{\bf k}}{\partial t} \, d{\bf k}
	&=\pi \int    \frac{
		\alpha^2
		}{\left(  \Lambda +p_1^2  \right)^2}    \delta(-2{\bf p}_1\cdot{\bf p}_2)
			n_2^2  n_k^2
			({\bf p}_1\cdot 2{\bf k})^2
			\left(  \partial_\omega n_\omega^{-1}-\partial_{\omega_2} n_{\omega_2}^{-1}  \right)
			\left(  \partial_\omega \varphi_\omega-\partial_{\omega_2} \varphi_{\omega_2}  \right)
			d{\bf k}_1 \, d{\bf k}_2 \, d {\bf k},			
\end{align*}
where we have also used Eq.\ \eqref{eq:om_p1perpp2} in the argument of the frequency delta function, and exhausted the delta function of wavevectors by integrating out ${\bf k}_3$.

To constrain the integral to the resonant manifold, we fix ${\bf k}$ and ${\bf k}_2$ and change variables from ${\bf k}_1$ to $(p_1, \theta)$, where $\theta$ is the angle between ${\bf p}_1$ and ${\bf p}_2$,
see Fig.\ \ref{fig:angles}(b).
The volume element transforms as $d{\bf k}_1 = p_1 dp_1\, d\theta$, and we perform the $\theta$ integral as follows:
\begin{align*}
\int  (\ldots)  \delta(-2{\bf p}_1\cdot {\bf p}_2) p_1 \, d\theta \, dp_1
	= \int  (\ldots)  \frac{1}{p_2} \, dp_1,
\end{align*}
where we have taken into account the fact that ${\bf p}_1$ and ${\bf p}_2$ are orthogonal, see Appendix \ref{app:triangle}. 
Then, using the properties of the scalar triple product, we can write
${\bf k}  \cdot  {\bf p}_1 = \pm  {\bf k}  \cdot  (  {\bf e}_z  \times  {\bf p}_2  )  p_1/p_2 
								  = \pm  {\bf e}_z  \cdot  ({\bf p}_2 \times  {\bf k}  )  p_1/p_2$.
Thus, we have
\begin{align}
\int \varphi_{\bf k}\frac{\partial n_{\bf k}}{\partial t}  \, d{\bf k}
	= 4 \pi    \int    \frac{
		\alpha^2
		p_1^2}{\left(  \Lambda +p_1^2  \right)^2}   
			 \frac{|{\bf p}_2 \times {\bf  k}|^2}{p_2^3}    n_2^2  n_k^2
			\left(  \partial_\omega n_\omega^{-1}-\partial_{\omega_2} n_{\omega_2}^{-1}  \right)
			\left(  \partial_\omega \varphi_\omega-\partial_{\omega_2} \varphi_{\omega_2}  \right)
			dp_1 \,  d{\bf k}_2 \, d{\bf k}.    						\label{eq:int_fndot_2}
\end{align} 

The dependence on $p_1$ can be factored into the reduced interaction coefficient, which can be calculated exactly,
\begin{align}\label{eq:S_Lambda}
S_\Lambda  =  4\pi  \int_0^\infty  \frac{
	\alpha^2
		p_1^2}{\left(  \Lambda +p_1^2  \right)^2} \, dp_1 
				   =  \frac{\pi^2 		  
					\alpha^2
					}{\sqrt{\Lambda}}.
\end{align}
This last step highlights the important feature of the SHE in this analysis: it is in this step that we have used the peaked nature of the interaction coefficient to reduce it to the coefficient $S_\Lambda$ analytically. To our knowledge, the SHE 
constitute the first system analysed in wave turbulence theory whose interaction coefficient can be reduced in this way.

To express the ${\bf p}_2$  dependence of \eqref{eq:int_fndot_2} in terms of ${\bf k}_2$ and ${\bf k}$, we denote the angle between vectors ${\bf k}$ and ${\bf k}_2$ by $\phi$
(see Fig.\ \ref{fig:angles}(a)),
note that $|{\bf p}_2 \times {\bf k} |^2 = |{\bf k}_2 \times {\bf k} |^2 = k_2^2k^2\sin^2(\phi)$, and use basic trigonometry to re-express $p_2$, giving
\begin{align*}
\int \varphi_{\bf k}  \frac{\partial n_{\bf k}}{\partial t} \, d{\bf k}
		= 
				S_\Lambda    \int    \frac{k_2^2 k^2 \sin^2(\phi)}{(k_2^2  -  2 k_2 k \cos(\phi)  +  k^2)^{3/2}}    n_2^2  n_k^2
				\left(  \partial_\omega n_\omega^{-1}-\partial_{\omega_2} n_{\omega_2}^{-1}  \right)
				\left(  \partial_\omega \varphi_\omega-\partial_{\omega_2} \varphi_{\omega_2}  \right)   
				d{\bf k}_2  \, d{\bf k} .
\end{align*}
The ${\bf k} \leftrightarrow {\bf k}_2$ symmetry of the integrand allows us to replace $(\ldots)\left( \partial_\omega \varphi_\omega-\partial_{\omega_2} \varphi_{\omega_2} \right) \to 2(\ldots)\left( \partial_\omega \varphi_\omega\right)$.

The next step is to move to frequency space by writing the integrations over ${\bf k}$ and ${\bf k}_2$ in polar form, so that 
$ d{\bf k} = (1/2) d \chi \, d\omega$ and $d{\bf k}_2 = (1/2)  d\phi \, d\omega_2$. Here $\chi$ is the polar angle of wavevector ${\bf k}$
(see Fig.\ \ref{fig:angles}(a)),
which we integrate out immediately and cancel from both sides. We obtain
\begin{align}
\int \varphi_\omega  \frac{\partial  n_\omega}{\partial t}  \, d\omega
		&= S_\Lambda    \int    \frac{\omega_2 \omega \sin^2(\phi)}{(\omega_2  -  2 \sqrt{\omega_2 \omega} \cos(\phi)  +  \omega)^{3/2}}    
				n_{\omega_2}^2  n_\omega^2
				\left(  \partial_\omega n_\omega^{-1}-\partial_{\omega_2} n_{\omega_2}^{-1}  \right)
				\left(  \partial_\omega \varphi_\omega  \right)   
				d\phi \,  d\omega_2 \, d\omega .
\end{align}

Finally, we integrate by parts with respect to $\omega$ to isolate the test function $\varphi_\omega$ on both sides, and use the fact that $\varphi_\omega$ is arbitrary, to obtain the SLAM for the SHE:
\begin{subequations}\label{eq:SLAM}
\begin{align} \label{eq:SLAM_kinetic}
\frac{\partial n_\omega}{\partial t}  = - \frac{1}{\pi} \frac{\partial Q}{\partial \omega},
\end{align}
where
\begin{align} \label{eq:SLAM_Q}
Q = \pi S_\Lambda  \int 
				f \! \left( \! \sqrt{\frac{\omega_2}{\omega}}  \right)
				 \frac{\omega_2}{\sqrt{\omega}} \,
				 n_{\omega_2}^2  n_\omega^2
				\left(  \partial_\omega n_\omega^{-1}-\partial_{\omega_2} n_{\omega_2}^{-1}  \right) 
				d\omega_2 
\end{align}
is the waveaction flux flowing through $\omega$ in frequency space (or circle of radius $k=\sqrt{\omega}$ in wavevector space; the factor of $\pi$ arises from the transformation between the two spaces). 
	In Eq.\ \eqref{eq:SLAM_Q} the function $f(s)$ is defined as
\begin{align}
\label{eq:SLAM_f}
f(s)  =  \int_0^{2\pi}  \frac{\sin^2(\phi)}{(1-2s\cos(\phi)+s^2)^{3/2}}\, d\phi .
\end{align}
\end{subequations}
In Appendix \ref{app:f_properties} we note some properties of $f(s)$.

The SLAM, defined by Eqs.\ \eqref{eq:SLAM}, is the main result of the present paper. 
The SLAM must be supplemented with initial and boundary conditions to produce a well-posed problem. 
To retain consistency with the semilocal approximation, the initial condition can be any arbitrary function supported on $\omega \gg \Lambda$, 
and whose characteristic scale of variation $\partial \omega / \partial( \ln n_\omega) \gg \Lambda$. Respectively, these conditions ensure that $W^{12}_{3{\bf k}}$ decays sufficiently rapidly that superlocality is satisfied, and that the Taylor expansion in Eq.\ \eqref{eq:Taylorexpns} can be truncated.
As mentioned in section \ref{subsubsec:fjortoft}, we conjecture that the steady-state dual cascade spectra in the forced-dissipated problem will be independent of initial conditions.

As for the boundary conditions, if one were concerned with a solution on the whole real line then physicality requires that the spectrum and fluxes vanish as $\omega\to\infty$. The situation as $\omega\to 0$ is quite involved, as one expects solutions that exhibit finite-time blowup there, associated with condensation \cite{thalabard2021inverse}. 
At this point not only does the support of the spectrum violate the $\omega \gg \Lambda$ condition, but the system becomes strongly nonlinear, meaning that the WKE (and hence the SLAM) no longer describes the dynamics.
However, when examining the dual cascade spectra, we assume that dissipation forces the spectrum to vanish at $\omega_{d+}$ and $\omega_{d-}$. 
We take these as boundary conditions of the solution in sections \ref{sec:fluxdirections} and \ref{sec:inverse_cascade} when we characterise these spectra.


\subsection{Conservation of invariants in the SLAM}\label{subsubsec:conservation}	
	
To show that the original invariants of the WKE continue to be conserved in the SLAM, we first note that \eqref{eq:SLAM_kinetic} is a continuity equation for waveaction, and so $N$ is manifestly conserved. 
Secondly, we note that the energy density is $\omega n_\omega$, and so the continuity equation for energy is 
\begin{align}\label{eq:E_conty}
\frac{\partial(\omega n_\omega)}{\partial t}  = - \frac{1}{\pi}\frac{\partial P}{\partial\omega}
\end{align}
where $P$ is the energy flux. 
	Together with Eq.\ \eqref{eq:SLAM_kinetic}, this gives $\partial_\omega P = \omega \partial_\omega Q$. Integrating from $0$ to $\omega$, we obtain 
\begin{align}\label{eq:PofQ}
P(\omega) = \omega Q(\omega) - \int_0^\omega Q(\tilde{\omega}) \, d\tilde{\omega}.
\end{align}
Integrating Eq.\ \eqref{eq:E_conty} over all $\omega$ gives $\partial_t E = -[P(\infty)-P(0)]/\pi$. Using Eq.\ \eqref{eq:PofQ}, and assuming that the particle flux decays fast enough at large and small $\omega$, so that $\omega Q(\omega)|_{\infty} = \omega Q(\omega)|_0 = 0$, we obtain
\begin{align}
\frac{\partial E}{\partial t} 
 			 =  S_\Lambda  \int_0^\infty \! \int_0^\infty  
					f \! \left( \! \sqrt{\frac{\omega_2}{\tilde{\omega}}}  \right)
					\frac{\omega_2}{\sqrt{\tilde{\omega}}} \,
				 	n_{\omega_2}^2  n_{\tilde{\omega}}^2
					\left(  \partial_{\tilde{\omega}} n_{\tilde{\omega}}^{-1}-\partial_{\omega_2} n_{\omega_2}^{-1}  \right) 
				d\omega_2 d\tilde{\omega} .			\label{eq:Edot}
\end{align}
Now we observe that by Eq.\ \eqref{eq:f_oneover_s_app}, the factor $ (\omega_2/ \! \sqrt{\tilde{\omega}})  \,  f  ( \! \sqrt{\omega_2 /\tilde{\omega}}  )$ is symmetric under $\tilde{\omega} \leftrightarrow \omega_2$. 
This leaves the integrand on the right-hand side of Eq.\ \eqref{eq:Edot} antisymmetric under exchange of the integration variables, and so we must have $\partial_t E=0$. 

Therefore, in a closed system the SLAM preserves the same quadratic invariants as the WKE from which it is derived.
The rest of this paper is devoted to obtaining solutions of the SLAM, particularly the solutions that realise the dual cascade of invariants that is predicted by the Fj{\o}rtoft argument in a forced-dissipated system.

\section{Stationary solutions of the SLAM}
\label{sec:stationarysols}

In this section we show that the usual stationary solutions of the WKE---the equilibrium RJ spectrum, and the KZ cascade spectra---are stationary solutions of the SLAM \eqref{eq:SLAM}. 
We are particularly interested in spectra that are self-similar, i.e.\ of power-law form $n_\omega = C\omega^{-x}$, where $C$ is a constant that is positive for physical spectra.

\subsection{Thermodynamic equilibrium (RJ) spectrum}

The RJ spectrum describes the state of thermodynamic equilibrium where a linear combination of the integral invariants is partitioned equally over ${\bf k}$-space:
\begin{align}\label{eq:RJ}
n_\omega = \frac{T}{\mu + \omega}  \quad  \text{(equipartition of $\mu N+E$)},
\end{align}
where the thermodynamic potentials are the temperature $T$ and chemical potential $\mu$ (both constants).
This spectrum is a stationary solution of the SLAM because the bracket $\left( \partial_\omega n_\omega^{-1}-\partial_{\omega_2} n_{\omega_2}^{-1}\right)$ in Eq.\ \eqref{eq:SLAM_Q} vanishes when Eq.\ \eqref{eq:RJ} is substituted.
	
The RJ spectrum has the asymptotic limits 
\begin{align}
\begin{aligned}
\label{eq:RJ_asymp}
n_\omega &\propto \omega^0      \quad  \text{(equipartition of $N$)}, \\
n_\omega &\propto \omega^{-1}  \quad  \text{(equipartition of $E$)},
\end{aligned}
\end{align}
which are self-similar spectra with spectral indices $x=0$ and $x=1$, respectively.

\subsection{Stationary nonequilibrium cascade (KZ) spectra}
\label{subsec:KZspectra}

As mentioned in Sec.\ \ref{sec:intro}, in many systems one can find stationary solutions of the WKE that are of power-law form, and which describe the constant flux of invariants via a self-similar, scale-by-scale cascade.
These are the KZ cascade spectra, and they are the first candidate for the spectra that realise the dual cascade predicted by the Fj{\o}rtoft argument
for forced-dissipated systems.
When the KZ spectra are physically relevant, the flux of each dynamical invariant will be described by its own KZ spectrum, and on that spectrum the flux of all other dynamical invariants will be zero.

To find the KZ spectra, we first substitute $n_\omega = C\omega^{-x}$ into Eq.\ \eqref{eq:SLAM_Q}, giving for the waveaction flux
\begin{align}\label{eq:Q_powerlaw}
Q      = \pi S_\Lambda C^3x     \int  
							f\! \left( \! \sqrt{\frac{\omega_2}{\omega}} \right)
							\frac{\omega_2}{\sqrt{\omega}} \,
							\omega_2^{-2x}\omega^{-2x}
							(\omega^{x-1}-\omega_2^{x-1}) \, d\omega_2 .
\end{align}

\subsubsection{Energy cascade spectrum}
\label{subsubsec:Ecascade}

In the wave turbulence literature, KZ spectra are frequently found by making non-identity transformations of the collision integral that allow one to read off spectral indices $x$ that make the integrand of the transformed collision integral vanish. This technique is known as the Zakharov-Kraichnan transform \cite{zakharov1965weak}. 

We now adapt this method to the SLAM in order to find the KZ energy cascade spectrum. We split the right-hand side of Eq.\ \eqref{eq:Q_powerlaw} into two halves. In the second half, we substitute $\omega_2 = \omega^2/\tilde{\omega}_2$. 
	Using Eq.\ \eqref{eq:f_oneover_s_app}, and dropping tildes immediately, we obtain
\begin{align*}
Q    =    \frac{\pi S_\Lambda C^3}{2}  x  \int    
							\left[  1-\left( \frac{\omega_2}{\omega} \right)^y  \right]
							f  \!  \left( \! \sqrt{\frac{\omega_2}{\omega}} \right)
							\frac{\omega_2}{\sqrt{\omega}} \,
							\omega_2^{-2x}\omega^{-2x} 
							(\omega^{x-1}-\omega_2^{x-1}) \, d\omega_2 , 
\end{align*}
with $y= 3x-3/2$. Choosing the spectral index $x=1/2$ leads to a vanishing waveaction flux $Q$, suggesting that this represents the KZ energy cascade spectrum.

To see that this is indeed the case, we extract the overall $\omega$ dependence in Eq.\ \eqref{eq:Q_powerlaw}, leaving a reduced, dimensionless collision integral $I(x)$, as follows:
\begin{align}\label{eq:Q_I}
Q(\omega) = 2\pi S_\Lambda C^3 \omega ^{(1-6x)/2} I(x),
\qquad\text{where}\qquad
I(x) =  x \int_0^\infty f(s) s^{3-4x} (1-s^{2x-2}) \, ds ,
\end{align}
and $s=\sqrt{\omega_2/\omega}$.
	Substituting Eq.\ \eqref{eq:Q_I} into Eq.\ \eqref{eq:PofQ} gives for the energy flux
\begin{align}\label{eq:P_I}
P(\omega) = 2\pi S_\Lambda C^3 \frac{1-6x}{3-6x} \omega^{(3-6x)/2} I(x).
\end{align}

Setting $x=1/2$ in Eq.\ \eqref{eq:Q_I} reproduces the result that $Q(\omega)=0$, because $I(1/2) \propto \int_0^\infty \! sf(s) \, ds - \int_0^\infty \! f(s) \,ds$, which vanishes by the $s\to1/s$ symmetry in Eq.\ \eqref{eq:f_oneover_s_app} (note that the transformation $s\to 1/s$ is exactly equivalent to making the Zakharov-Kraichnan transform).

When $x=1/2$,  Eq.\ \eqref{eq:P_I} gives $P(\omega)=0/0$. To resolve this indeterminacy we use L'H{\^o}pital's rule, obtaining
 \begin{align*} 
 P(\omega)
			 =\frac{2\pi S_\Lambda C^3}{3} I'(1/2) 
 			= \frac{2\pi S_\Lambda C^3}{3} \left(3 \int_0^\infty f(s) 
			\ln
			 (s) \, ds \right)  
 			= -4.85432 \left( \pi S_\Lambda C^3 \right),
 \end{align*}
 where the prime denotes differentiation with respect to $x$, and we have again used Eq.\ \eqref{eq:f_oneover_s_app}.
 
Thus, when $x=1/2$ the energy flux $P$ is a constant, independent of $\omega$, while the waveaction flux $Q$ vanishes, indicating that this is indeed the KZ energy cascade spectrum. However, we note that on this spectrum the sign of $P$ is negative, which is opposite to the sign predicted by the Fj{\o}rtoft argument. We elaborate on this in Sec.\ \ref{sec:fluxdirections}.

\subsubsection{Waveaction cascade spectrum}
\label{subsubsec:Ncascade}

To determine the KZ spectrum for the cascade of waveaction, we put $x=1/6$ in Eq.\ \eqref{eq:Q_I}, obtaining $Q(\omega) = 2\pi S_\Lambda C^3 I(1/6)$, which is independent of $\omega$. Likewise, Eq.\ \eqref{eq:P_I} gives $P(\omega)=0\times I(1/6)$. This would satisfy the requirements to be the KZ waveaction cascade spectrum if $I(1/6)$ converged. However, from the second equation in \eqref{eq:Q_I} we see that 
\begin{align*}
I(1/6) = \frac{1}{6}\int_0^\infty f(s)(s^{7/3} - s^{2/3})\, ds .
\end{align*}
Noting the asymptotic behaviour of $f(s)$ from Eq.\ \eqref{eq:f(infty)_asymp_app}, we see that $I(1/6)$ diverges as $s\to\infty$. 

Therefore, even though the power-law spectrum with $x=1/6$ superficially gives the correct properties for the KZ waveaction cascade spectrum, we must rule it out because the collision integral is divergent on that spectrum.

\subsubsection{Summary of KZ spectra}

To summarise, the formal KZ cascade spectra that are our first candidates for realising the dual cascade are
\begin{align}
\begin{aligned}
\label{eq:KZ}
n_\omega &\propto \omega^{-1/2} 		 \quad  \text{(KZ spectrum: cascade of $E$)}, \\
n_\omega &\propto \omega^{-1/6}      \quad  \text{(KZ spectrum: cascade of $N$)}.
\end{aligned}
\end{align}
However, both spectra suffer pathologies: on the first spectrum the flux of energy is in the wrong direction, and the second spectrum causes the collision integral to diverge. We must therefore rule them out, and seek other spectra on which the dual cascade can be supported.

These pathologies notwithstanding, it is still worth noting that the original interaction coefficient $W^{12}_{3{\bf k}}$ is not a homogeneous function of the four wavevectors, i.e.\ it possesses no obvious properties that would lead to a self-similar scaling behaviour.
Nevertheless, the semilocality property of $W^{12}_{3{\bf k}}$ allows us to integrate out its non-homogeneous part, giving the constant coefficient $S_\Lambda$. The resulting equation, the SLAM, is self-similar. However, unlike KZ spectra, the relevant solutions of the SLAM that manifest the dual cascade do not turn out to be self-similar themselves, as we demonstrate in the following sections.

\subsection{Interpretation of divergent spectra}
\label{subsec:divergence_interpretation}
As mentioned above, the divergence of the collision integral at a certain scale causes us to rule out a prospective KZ spectrum. However, we expect the true solution to retain some characteristics indicated by this divergence. Namely, waves of a scale that approaches the divergent scale will be increasingly dominant in every tetrad of interacting waves in which they participate. In other words, wave interactions at every scale will be mediated by waves whose scale approaches the divergent scale. In this situation, the true solution is termed a nonlocal flux spectrum, as opposed to the spectrally-local cascades that are described by physically-realisable KZ solutions. 

In the specific case here, the divergence of the KZ waveaction cascade spectrum as $s\to\infty$ implies that the true waveaction flux spectrum is nonlocal, dominated by interactions at $\omega_2 \gg \omega$.	
By contrast, the fact that the KZ energy cascade spectrum gives convergence of the collision integral signals that the true cascade solution has local interactions. We need only resolve the matter of the cascade direction, which we do in Sec.\ \ref{sec:fluxdirections}.
	
In Appendix.\ \ref{app:locality} we present a full convergence study of the collision integral on general power-law spectra, allowing us to see the KZ spectra in their full context. The results of this convergence study are shown in Fig.\ \ref{fig:indices_fluxes_convergence}(b).

\section{Flux directions on power-law spectra}\label{sec:warm}
\label{sec:fluxdirections}

For the sake of completeness, we present in this section a general diagrammatic argument \cite{nazarenko2011book} that determines the directions of both the energy flux $P$ and the waveaction flux $Q$, on all power-law spectra $n_\omega = C \omega^{-x}$ (with $C>0$). In order to present the argument, we neglect for a moment the divergence of the collision integral on the KZ waveaction cascade spectrum.
	Recall that if the sign of a flux is positive (negative), the invariant flows towards large (small) $\omega$. 
		
	It is natural to assume that for very sharply peaked spectra, the resulting fluxes will flatten the spectra out. Thus, for $x\to\infty$ (spectrum sharply peaked around $\omega\approx 0$), we expect the associated fluxes to be strongly positive.
	Likewise for $x\to -\infty$ (spectrum sharply rising), the fluxes will be strongly negative.
	
\begin{figure}[t]
\begin{center}
\includegraphics[width=0.75\columnwidth]{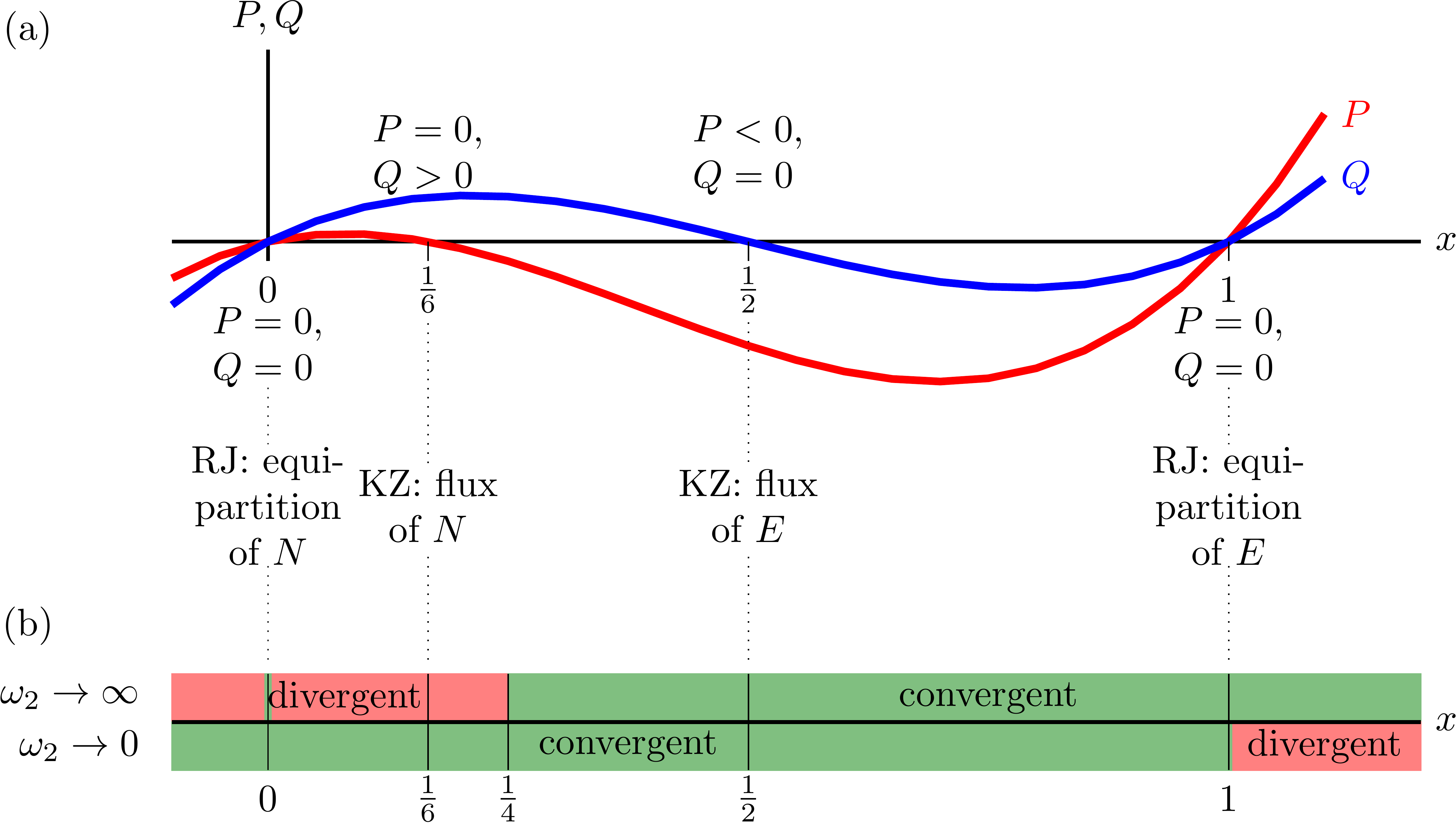}
\caption{\label{fig:indices_fluxes_convergence}
	(a) Sketch of the energy flux $P$ and waveaction flux $Q$ dependence on spectral index $x$, where $n_\omega = C \omega^{-x}$. The signs of the fluxes are determined by the relative ordering of the RJ and KZ spectra (Eqs.\ \eqref{eq:RJ_asymp} and \eqref{eq:KZ} respectively), where one or both of the fluxes is zero, and the behaviour at large and small $x$. The qualitative behaviour of the fluxes in between the zeros follows by continuity, see Sec.\ \ref{sec:fluxdirections}.
	(b) Convergence (green) or divergence (red) of Eq.\ \eqref{eq:SLAM_Q} with respect to spectral index $x$. Above the $x$ axis refers to $\omega_2$ in the ultraviolet range, and below the $x$ axis refers to $\omega_2$ in the infrared range, see Appendix \ref{app:locality}. (Convergence is unconditional exactly on the thermodynamic spectra $x=0,1$. This is indicated by the narrow green strips around these two spectra.)
	}
\end{center}
\end{figure}

	In between these two, the fluxes will both be zero on each of the thermodynamic spectra $x=0,1$. 
	As for the KZ spectra, by construction the KZ energy cascade spectrum is for the \emph{pure} flux of energy, with no waveaction flux.
	Likewise, the energy flux is zero on the KZ spectrum for a pure waveaction cascade. 
	In our case we respectively have $Q=0$ for $x=1/2$, and $P=0$ for $x=1/6$ (were the latter to give a convergent collision integral).
	Assuming that the fluxes vary continuously with spectral index $x$ forces them to behave qualitatively as shown in Fig.\ \ref{fig:indices_fluxes_convergence}(a).

We see that the ordering of the zero crossings forces $P$ to be negative on the KZ energy cascade spectrum, as found in Sec.\ \ref{subsubsec:Ecascade}, and also forces $Q$ to be positive on the KZ waveaction cascade spectrum. These are both in direct contradiction to the conclusion of the Fj{\o}rtoft argument, which is that $P$ must be positive and $Q$ negative on the stationary spectra that realise the 
direct and inverse portions of the dual cascade,
see Sec.\ \ref{subsubsec:fjortoft}. 
If the respective 
flux-carrying solutions	
are to be realised by the KZ spectra, the only way to reconcile the two arguments is for the KZ spectra to have non-positive (i.e.\ negative or even complex) prefactor constants $C$, which is clearly unphysical.
Therefore, we conclude once more that the KZ spectra found in Sec.\ \ref{sec:stationarysols} cannot realise the dual cascade in any physically-relevant scenario.

As the KZ spectrum must be ruled out, the true solution to realise a steady-state cascade must be related to the other stationary solution: the RJ spectrum \cite{dyachenko1992optical}.
Indeed, experience with other wave turbulence systems suggests that the true cascade solution is an RJ spectrum with small deviations that are nonetheless responsible for carrying the entire flux, see Eq.\ \eqref{eq:direct_spectrum}.
Such solutions are termed warm cascade spectra \cite{dyachenko1992optical, proment2012warm, skipp2020wave}.

We therefore hypothesise that the flux-carrying spectra that realise the dual cascade are warm spectra in both the direct and inverse inertial ranges. 		
However, anticipating the results of Sec.\ \ref{sec:inverse_cascade}, we will conclude that the inverse cascade of $N$ is not only nonlocal in character, but is also realised by a warm spectrum with negative thermodynamic potentials $T$ and $\mu$. 

\begin{figure}[t]
\begin{center}
\includegraphics[width=0.75\columnwidth]{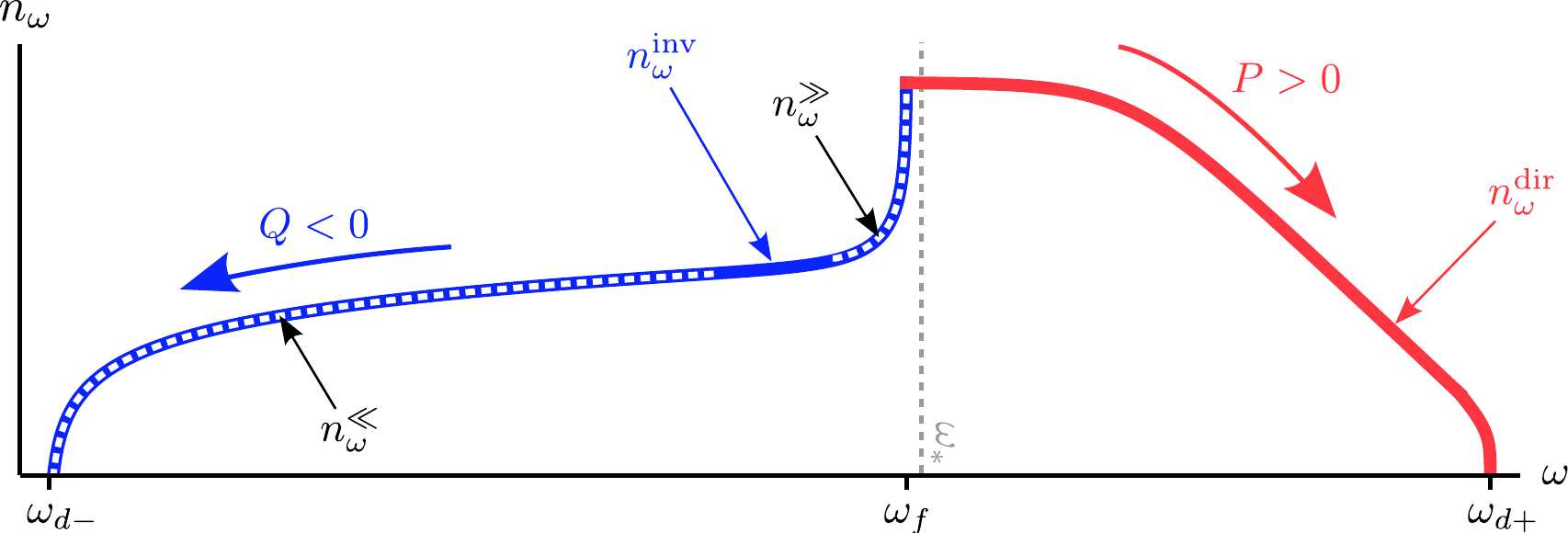}
\caption{\label{fig:Dual_cascade}
Qualitative sketch of the steady-state dual cascade predicted by the SLAM, which realises the prediction of the Fj{\o}rtoft argument. Forcing at $\omega_f$ injects waveaction and energy into the system. The negative waveaction flux $Q$ is realised by the 
nonlocal
inverse cascade spectrum $n_\omega^\text{inv}$ (blue), which terminates at the scale $\omega_{d-}$ where the majority of the waveaction is dissipated. The positive energy flux $P$ is realised by the
local
direct cascade spectrum $n_\omega^\text{dir}$ (red), which terminates at $\omega_{d+}$ where most of the energy is dissipated. The asymptotic solutions $n_\omega^\ll$ and $n_\omega^\gg$ are overlaid in white dashes. 
(Note that the inverse and direct cascade spectra meet at $\omega_f$, which is strictly less than, but of the same order as, $\omega_*$. In a realistic system, the break in gradient at $\omega_f$ will be regularised by the specific forcing protocol, which we do not attempt to show here.)
}
\end{center}
\end{figure}

By contrast, the convergence of the KZ energy cascade spectrum found in Sec.\ \ref{subsubsec:Ecascade}, and the discussion of Sec.\ \ref{subsec:divergence_interpretation}, indicate that the true direct cascade of $E$ is local. We therefore expect the direct cascade to be warm, with positive $T$ and $\mu$, and spectrum
\begin{equation}
\label{eq:direct_spectrum}
n_\omega^\text{dir} = \frac{T}{\mu + \omega +\Delta(\omega)}.
\end{equation}
Here, $\Delta(\omega)$ is the deviation from the RJ spectrum, which remains small in the inertial range, far from the forcing and dissipation scales. At the end of the inertial range $\Delta(\omega)$ becomes large, until the spectrum terminates at the dissipation scale $\omega_{d+}$. 
We sketch the warm direct cascade $n_\omega^\text{dir}$ qualitatively in red in Fig.\ \ref{fig:Dual_cascade}.

It is a prediction from the superlocal DAM that the warm spectrum terminates in a logarithmic compact front, and that the temperature of the cascade spectrum $T$ is determined by the energy flux $P$, and small-scale dissipation range $\omega_{d+}$ \cite{skipp2020wave}.
We leave it to future work, reinforced by numerical simulations, to examine these relations for the direct warm cascade realised by the SLAM.


\section{Nonlocal inverse cascade solution}\label{sec:inverse_cascade}

In this section, we seek the stationary solution of the SLAM that realises a constant inverse flux of waveaction, and that is nonlocal in the sense suggested by the divergence of the corresponding KZ spectrum, see Sec.\ \ref{subsec:divergence_interpretation}. 
We also seek to parameterise the solution in terms of quantities that we can control externally, for example in simulations. These will turn out to be the flux $Q$, the forcing and dissipation scales $\omega_f$ and $\omega_{d-}$, and the temperature of the inverse warm cascade $T$.

We set $Q$ to be negative in Eq.\ \eqref{eq:SLAM_Q} to specify an inverse flux, and substitute  $f(s)=\pi/s^3$, its $\omega_2 \gg \omega$ limit.
	Equation \eqref{eq:SLAM_Q} becomes
\begin{align}\label{eq:nonloc_diff}
\frac{\partial n_\omega}{\partial \omega}    =    \frac{\hat{Q}}{A\omega}   +   \frac{B}{A}n_\omega^2
\end{align}
where $\hat{Q}=-Q/\pi^2 S_\Lambda >0$ and the integrals over $\omega_2$ are absorbed into the constants
\begin{align}\label{eq:nonloc_A_B}
A = \int \frac{n_{\omega_2}^2}{\sqrt{\omega_2}} \, d\omega_2 
\quad \text{and} \quad
B = \int \frac{1}{\sqrt{\omega_2}}\frac{\partial n_{\omega_2}}{\partial \omega_2} \, d\omega_2.
\end{align}
Manifestly $A>0$. Self-consistency of the asymptotic solutions of Eq.\ \eqref{eq:nonloc_diff} demands that $B>0$ also (see discussion after Eq.\ \eqref{eq:nonloc_gg} below).

\subsection{Nonlocal inverse cascade: asymptotics}\label{subsec:nonloc_asymp}

First, we examine the asymptotics of Eq.\ \eqref{eq:nonloc_diff} to extract key characteristics of the full solution. We denote the frequency at which the two terms on the right-hand side are equal by $\omega_s$. 

	For $\omega \ll \omega_s$ the first term on the right-hand side of Eq.\ \eqref{eq:nonloc_diff} dominates. The solution to the resulting asymptotic equation is 
\begin{align}\label{eq:nonloc_ll}
n_\omega^\ll     =     \frac{\hat{Q}}{A}   
\ln
\left(  \frac{\omega}{\omega_{d-}}  \right).
\end{align}
Here we have written the constant of integration as the frequency at which the solution $n_\omega^\ll$ vanishes, and interpreted it as the dissipation scale $\omega_{d-}$. 
	Note that the solution finds a vanishing point naturally, without specifying a dissipation mechanism. This is in common with warm solutions of superlocal DAMs that contain compact fronts at which the solution vanishes logarithmically, see e.g.\ \cite{proment2012warm, skipp2020wave}.
	(We expect that if dissipation is not provided, so that the flux is drained from the system by the time it reaches $\omega_{d-}$, the spectrum would grow in this vicinity so that the situation would not be time-independent. Eventually the nonlinearity would become strong here, so that the wave turbulence assumptions would become violated.)

	For $\omega \gg \omega_s$ the second term on the right-hand side of Eq.\ \eqref{eq:nonloc_diff} is dominant, and we have the asymptotic solution
\begin{align}\label{eq:nonloc_gg}
n_\omega^\gg     =     \frac{A/B}{\omega_*   -   \omega}.
\end{align}
Here the constant of integration appears as $\omega_*$, the frequency at which the solution becomes singular. By hypothesis, $\omega_*$ is greater than any $\omega$ in the inverse cascade range. The integral of $n_\omega^\gg$ is weakly (logarithmically) divergent as $\omega\to\omega_*$, which is consistent with the assumption of a nonlocal solution that is dominated by interactions with $\omega_2\gg\omega$.
	Obviously, in any realistic scenario the solution cannot continue up to $\omega_*$. We therefore cut the solution off at $\omega_f$ where $\omega_{d-} \ll \omega_f < \omega_*$. This cutoff represents the end of the inverse cascade inertial range; in a forced-dissipated setup this is none other than the forcing scale. By choosing $\omega_f$ in the vicinity of $\omega_*$, so that $\omega_2$ can approach the singularity frequency $\omega_*$, we keep consistency with the nonlocality assumption $\omega_2\gg\omega$.

	If we define the temperature $T\coloneqq -A/B$ and chemical potential $\mu \coloneqq -\omega_*$, we also see that $n_\omega^\gg$ is actually a thermodynamic spectrum \eqref{eq:RJ} with negative $T$ and $\mu$. 
	The interpretation of RJ equilibria with negative thermodynamic potentials was given in 
Ref.\ \cite{skipp2021equilibria} for the case of three sign-definite invariants. For the present case with two invariants, these are exactly equilibria with spectra diverging at some nonzero $\mu$ (see appendix of Ref.\  \cite{skipp2021equilibria}).

	Note that, had we chosen $B<0$, we would have obtained the asymptotic solution $n_\omega^{B<0} =A/|B|(\omega+\omega_*)$. For $\omega_*<0$ this is negative in $0<\omega<|\omega_*|$, which is unphysical. 
	For $\omega_*\geq0$, if we substitute $n_\omega^{B<0}$ back into Eq.\ \eqref{eq:nonloc_diff}, the first term on the right-hand side dominates for all $\omega \geq 0$, which is inconsistent with the assumptions for deriving $n_\omega^{B<0}$. We therefore rule out the $B<0$ case.
	Had we chosen $B=0$, the full solution $n_\omega^\ll$ is the only solution, but then the second integral in \eqref{eq:nonloc_A_B} gives $B\neq0$. 
	Hence, we rule out $B=0$ as well, and therefore we must have $B>0$.

	Thus, the full solution of Eq.\ \eqref{eq:nonloc_diff} resembles an RJ spectrum for $\omega \gg \omega_{d-}$, Eq.\ \eqref{eq:nonloc_gg}, but has
a deviation that grows towards the infrared, and that terminates at $\omega = \omega_{d-}$ with a logarithmic compact front, Eq.\ \eqref{eq:nonloc_ll}.
This is exactly to say that it is a warm cascade spectrum, but with negative thermodynamic potentials.

\subsubsection{Determination of constants $A$ and $B$}
	
	The integrals in Eqs.\ \eqref{eq:nonloc_A_B} must be taken over the whole inverse cascade range, from $\omega_{d-}$ up to $\omega_f$. Since the inverse cascade spectrum is nonlocal, the dominant contributions to the integrals occurs at large $\omega$. 
	Using the asymptotic spectrum $n_\omega^\gg$, and evaluating Eqs.\ \eqref{eq:nonloc_A_B} at the upper limit $\omega_f$, we obtain, to leading order, $B^2/A = 1/\sqrt{\omega_f}(\omega_*-\omega_f)$. In terms of the temperature $T$ this gives
\begin{align}\label{eq:nonloc_A_B_param}
A = \frac{T^2}{\sqrt{\omega_f}(\omega_*-\omega_f)}
\quad\text{and}\quad
B = -\frac{T}{\sqrt{\omega_f}(\omega_*-\omega_f)},
\end{align}
i.e.\ we have expressed $A$ and $B$ in terms of $T, \omega_f$ and $\omega_*$.

\subsection{Nonlocal inverse cascade: full solution}\label{subsec:nonloc_full}

	Equation \eqref{eq:nonloc_diff} can be solved analytically by noting that it is a Ricatti equation. Using standard techniques \cite{polyanin2017handbookODEs}, its solution is found to be
\begin{align}\label{eq:nonloc_full}
n_\omega^\text{inv}      =     -\sqrt{\frac{\hat{Q}}{B \omega}}  \left(\frac{Y_0(K) + J_0(K)c}{Y_1(K) + J_1(K)c} \right),
\quad\text{with}\quad
K=\frac{2 \sqrt{B \hat{Q}\omega}}{A},
\end{align} 
where $J_n(K), Y_n(K)$ are $n$-th order Bessel functions of the first and second kinds respectively, and $c$ is the constant of integration.

	We can relate $c$ to the integration constants of the asymptotic solutions by noting that $\omega_{d-}$ corresponds to the first zero of the right-hand side of Eq.\ \eqref{eq:nonloc_full}. This will be at the first root of the numerator $Y_0(K) + J_0(K)c$. 
	Using the asymptotics of the Bessel functions for $K \ll 1$ gives, to leading order,
\begin{align}
\label{eq:nonloc_omegadminus_c}
	\omega_{d-} = \frac{A^2}{B\hat{Q}} e^{-2 \gamma -\pi c},
\end{align}
where $\gamma\approx 0.5772$ is the Euler-Mascheroni constant.
	Likewise, $\omega_*$ corresponds to the first root of $Y_1(K) + J_1(K)c$, the denominator of Eq.\ \eqref{eq:nonloc_full}. To leading order this gives 
\begin{align}
\label{eq:nonloc_omegastar_c}
	\frac{A^2}{B\hat{Q}\omega_*}  +  
	\ln
	\left( \frac{A^2}{B\hat{Q}\omega_*}  \right)
		= \pi c + 2 \gamma- 1.
\end{align}
This has solution $\omega_* = A^2/[B\hat{Q} \,W(e^{\pi c + 2 \gamma -1})]$, where $W(x)$ is the Lambert-W function. 

Firstly, we note that the first term on the right-hand side of Eq.\ \eqref{eq:nonloc_omegastar_c} is dominant as we send $c\to\infty$, while the left-hand side is greater than $A^2/(B\hat{Q}\omega_*)$. This gives $\omega_*^{-1}(c) = \mathcal{O}(c)$, whereas from Eq.\ \eqref{eq:nonloc_omegadminus_c}, $\omega_{d-} \to 0$ exponentially as $c\to\infty$. Thus, the ratio $\omega_{d-}/\omega_*\to 0$ as $c\to\infty$, and so by adjusting $A,B,$ and $\hat{Q}$ to set the overall scaling, we can make the inverse inertial range arbitrarily wide.

Furthermore, we can eliminate $c$ between Eqs.\ \eqref{eq:nonloc_omegadminus_c} and \eqref{eq:nonloc_omegastar_c}, and eliminate $A$ and $B$ using Eq.\ \eqref{eq:nonloc_A_B_param}, obtaining
\begin{align}
\label{eq:nonloc_omegastar}
	 \ln
	 \left(   \frac{  \omega_* }{ \omega_{d-}  }  \right) 
		= 1+ \frac{  T^3  }{  \hat{Q}  \sqrt{\omega_f}  (\omega_* - \omega_f) \omega_*  }.
\end{align}
Equation \eqref{eq:nonloc_omegastar} implicitly expresses $\omega_*$ in terms of the control parameters $(\hat{Q}, T, \omega_f, \omega_{d-})$.
Solving for $\omega_*$ (for example numerically, or to any desired accuracy by iteration), and substituting into Eq.\ \eqref{eq:nonloc_A_B_param} allows $A$ and $B$ to be written in terms of the same set of parameters.
	We can likewise express $c$ via Eq.\ \eqref{eq:nonloc_omegadminus_c}, and finally obtain the solution $n_\omega$ via Eq.\ \eqref{eq:nonloc_full}, in terms of the control parameters $(\hat{Q}, T, \omega_f, \omega_{d-})$. 

Note that, unlike the case for superlocal DAMs, we cannot close the set of control parameters by writing $T$ as a function of the flux and the forcing and dissipation scales. This is reminiscent of two-free-parameter stationary solutions of the Leith model \cite{connaughton2004warm}. (Closures could be provided by specific assumptions about the forcing or dissipation, for example that the forcing starts with a given flux and temperature, but these assumptions would not be universal.)

In Fig.\ \ref{fig:Dual_cascade} we sketch the qualitative behaviour of the nonlocal inverse waveaction cascade spectrum $n_\omega^\text{inv}$ in blue. We also show the asymptotic solutions $n_\omega^\ll$ and $n_\omega^\gg$ in white dashes, and the frequency $\omega_*$ where the inverse cascade spectrum becomes singular.

\section{Discussion and conclusion}
\label{sec:discussion_conclusion}

\subsection{Comparison with the NLSE and SNE limits}
\label{subsec:cf2020}	

	Before concluding, we make some remarks about the two limits of the SHE that were mentioned in Sec.\ \ref{sec:SHE}. The first is the NLSE limit, where we send $
	\alpha, 
	\Lambda \to \infty$, while $
	\alpha 
	/\Lambda \to \text{const.}$ 
	After rescaling $\psi$ we obtain Eq.\ \eqref{eq:NLS}. The second is the SNE limit, where we set $\Lambda=0$ to obtain Eqs.\ \eqref{eq:SNE}. 

	In both these limits, the interaction coefficient becomes a homogeneous function, in the sense that $W^{\mu{\bf k}_1, \mu{\bf k}_2}_{\mu{\bf k}_3, \mu{\bf k}}  =  \mu^\beta W^{{\bf k}_1, {\bf k}_2}_{{\bf k}_3, {\bf k}}$, with $\beta=0$ in the NLSE limit and $\beta=-2$ in the SNE limit. 
	This observation allowed us to heuristically construct superlocal DAMs of the NLSE and SNE in Ref.\ \cite{skipp2020wave}. 
	There, we used the DAMs to examine the respective KZ spectra, and found that in both cases the flux directions contradicted the Fj{\o}rtoft argument.
	We therefore proposed that the flux-carrying spectra were warm spectra in both the NLSE and SNE. 
	We are now in a position to revisit this work, in light of the rigorously-derived SLAM. 

	The first thing to note is that in the NLSE limit, the interaction coefficient becomes a constant across all wavevectors. 
	In particular, this no longer respects the semilocality property: pairs of wavevectors are no longer picked out by the sharp decay of the interaction coefficient. 
	We therefore cannot approximate the full WKE by the SLAM---to do so would neglect the majority of wave interactions, all of which are important in evolving the spectrum. 
	However, we can still use the DAM for qualitative understanding, e.g. the argument about flux directions and the prediction of warm cascades \cite{dyachenko1992optical, skipp2020wave}.

	By contrast, the SNE are a singular limit of the SHE. We noted in Sec.\ \ref{sec:SHE} that the SNE are ill-posed, and that their regularisation requires restoring $\Lambda \neq 0$, i.e.\ moving to the SHE. 
	To elaborate: in order to develop the wave turbulence theory and derive the WKE, one starts with a periodic system \cite{nazarenko2011book}, but in a periodic system, Eq.\ \eqref{eq:SNE_Pois} has no non-trivial solutions. 
	Once we derive the SLAM, this ill-posedness is revealed in Eq.\ \eqref{eq:S_Lambda}: setting $\Lambda=0$ sends $S_\Lambda \to \infty$, i.e.\ the SLAM diverges for every spectrum $n_\omega$. 
	This indicates that one can formally write down the kinetic equation of the SNE, but the collision integral becomes infinite when any two wavevectors become equal.
	 Likewise, one can obtain KZ spectra for the SNE based on dimensional arguments, but these spectra will be invalid because the collision integral will be divergent on these spectra.  Moreover, the KZ spectra will
change discontinuously when we regularise the kinetic equation by setting $\Lambda \neq 0$.
	This is indeed what we find when we compare the KZ spectra found in Ref.\ \cite{skipp2020wave} (namely $\omega^0$ for the KZ waveaction cascade spectrum, and $\omega^{-1/3}$ for the KZ energy cascade spectrum, for the 2D case) to Eqs.\ \eqref{eq:KZ}.

	Thus, we see that retaining $\Lambda \neq 0$ in the SHE is necessary in order to regularise the singular SNE limit.	
	This is a salutary lesson as it highlights the hidden pitfalls of such heuristic derivations of DAMs: their predictions are misleading if, as in our case, they do not respect essential properties of the original interaction coefficient. 
	We speculate that a similar derivation of a semilocal model might be applied to other examples in the literature,
e.g.\ in the theory of gravitational waves in Einstein's vacuum field model \cite{galtier2017turbulence}.

\subsection{Conclusion}
\label{subsec:conclusion}

	Starting from the wave kinetic equation of the SHE, we have rigorously derived a reduced kinetic equation, the SLAM, by exploiting the natural locality properties of the interaction coefficient. 
	We believe this to be the first such derivation of a reduced kinetic equation in which the locality assumption can be justified self-consistently. 

	Having derived the SLAM, we use it to obtain the stationary spectra that are responsible for realising the dual cascade of energy and waveaction that is predicted by the Fj{\o}rtoft argument. After deriving the formal KZ cascade spectra, and examining their flux directions and locality, we conclude that neither the direct cascade of energy nor inverse cascade of waveaction are realised by the respective KZ spectra.

Instead, we predict that the dual cascade is carried by warm spectra. This concurs with our examination of the limits of the SHE in Ref.\ \cite{skipp2020wave}, even though some of that was carried out in the SNE limit, which is, in fact, singular.
	Here, though, the SLAM allows us to refine our prediction about the character of the warm spectra. We predict that the direct energy cascade spectrum will have positive thermodynamic parameters, and that interactions will be between waves that are local in frequency. 
	By contrast, the inverse cascade of waveaction will be carried by a nonlocal spectrum, with interactions at every frequency $\omega$ being dominated by the spectrum near the forcing scale $\omega_f$.
	Accordingly, we derive a nonlocal, warm, inverse cascade spectrum, that is parameterised by a negative temperature and chemical potential.

Our results on the dual cascade were derived for the forced-dissipated 2D SHE, which is the setup that leads to the clearest manifestation of the cascades. 
Results on the inverse cascade may also apply to the case of turbulence that evolves freely from an initial condition
in a closed system,
due to the inverse cascade having finite capacity (the integral of the inverse cascade spectrum converges when we send $\omega_{d-}\to 0$). Experience with finite capacity KZ spectra shows that an initial condition fills out its respective inertial range in finite time, with the KZ spectrum establishing after an initial transient \cite{semikoz1995kinetics, semikoz1997condensation, galtier2000WT_MHD}. It remains to be tested whether this phenomenology carries over to the inverse cascade spectrum of the SHE. 
By contrast, the direct cascade has infinite capacity for energy (the integral defining energy diverges as $\omega_{d+}\to \infty$), and so it can absorb an arbitrary amount of energy that is sent into it, unless there is some small-scale cutoff that arrests the direct cascade e.g.\ the finite numerical resolution of a simulation. 
For such systems, the cascade spectrum typically does not form behind the front that propagates from an initial condition, unless continuously forced.
	 
The finite-capacity inverse cascade cannot absorb an arbitrary amount of waveaction. If no large-scale dissipation is provided, waveaction will arrive at the end of the inertial range and start to accumulate into coherent large-scale structures: condensates and solitons. 
The dynamics of these structures are attracting much interest, particularly in astrophysics, where they could represent galactic dark matter halos \cite{mocz2017galaxyI, hui2017ultralight, levkov2018gravitational, li2019numerical}, or their 1D and 2D analogues in 
nonlinear optics 
\cite{Bekenstein2015_OpticalNewtSchro, paredes2020optics, garnier2021coherent, garnier2021incoherent}. 
The dual cascade process is a universal mechanism whereby such large-scale structures emerge due to the interaction of weakly turbulent small-scale waves, at least in the initial transient phase where weak waves exist without any coherent structures present. 
	
Our results are therefore directly applicable to the 2D case, which may be accessible in optical experiments.
Indeed, experiments modelled by the 2D SHE have already been conducted, using thermo-optic crystals as the nonlinear medium, particularly in the context of tabletop analogues of dark matter haloes or boson stars \cite{Bekenstein2015_OpticalNewtSchro, Faccio2016_OpticalNewtSchro, Vocke2018rotating}. 
1D liquid crystal experiments were carried out specifically looking at the wave turbulence of the SHE \cite{Laurie2012_1DOpticalWT}, and it would be feasible to extend this to a 2D experiment. 


The next step in this work is an extensive numerical comparison of the SHE, its WKE, and the SLAM, in both the freely-evolving and forced-dissipated case. This is currently in progress by the authors and will be published separately at a later date.  Such a comparison was recently carried out for the NLSE and its WKE \cite{zhu2022testing, zhu2023direct}.
Furthermore, we envisage that one could derive a similar SLAM for the 3D SHE, which will be applicable to the turbulent formation of galactic dark matter halos, and the 1D SHE, relevant to optical experiments such as those carried out in \cite{bortolozzo2009optical}. The same methodology should carry over to those cases, with the technical subtlety in the 1D case, the leading-order wave process is 6-wave, rather than the 4-wave case in 2D and 3D.

\section{Acknowledgements}

This work was supported by the European Union's Horizon 2020 research and innovation programme under the Marie Sk\l odowska-Curie grant agreement No. 823937 for the RISE project HALT, 
and by the Simons Foundation Collaboration grant Wave Turbulence (Award ID 651471). 
	J.L. and J.S. are supported by the Leverhulme Trust Project Grant RPG-2021-014.

\appendix

\section{Wavevectors relative to ${\bf k}$ form a right-angle triangle}
\label{app:triangle}

Waves on the resonant manifold are constrained to have a particular geometric relation.
	The frequency and wavevector resonance conditions \eqref{eq:resonance} give
\begin{align}
\omega^{12}_{3{\bf k}}   
		&=  ({\bf k}_1-{\bf k})\cdot({\bf k}_1+{\bf k}) + ({\bf k}_2-{\bf k}_3)\cdot({\bf k}_2+{\bf k}_3)		\nonumber \\
		&=  2({\bf k}_1-{\bf k})\cdot({\bf k}-{\bf k}_2)=0.  \label{eq:om_p1perpp2}
\end{align}
	Recalling the definition ${\bf p}_i \coloneqq {\bf k}_i-{\bf k}$, we thus have that ${\bf p}_1$ is orthogonal to ${\bf p}_2$. A similar calculation gives
the Pythagorean relation
$p_1^2+p_2^2=p_3^2$. 
We conclude that on the resonant manifold, ${\bf p}_1,{\bf p}_2,{\bf p}_3$ form a right-angle triangle,
see Fig.\ \ref{fig:angles}(b).

\section{Properties of $f(s)$}
\label{app:f_properties}

In this appendix we examine the function
\begin{align} \label{eq:f(s)_app}
f(s)    =    \int_0^{2\pi}     \frac{  \sin^2(\phi)  }{  (1 - 2s\cos(\phi) + s^{2})^{3/2}  }  \,  d\phi ,
\end{align}
where in our application to the SLAM \eqref{eq:SLAM}, $s=\sqrt{\omega_2/\omega}$ and so $0 \leq s < \infty$.

\subsection{Writing $f(s)$ in terms of complete elliptic integrals}

Writing Eq.\ \eqref{eq:f(s)_app} as a derivative, integrating by parts, using symmetry under $\phi \to 2\pi-\phi$, and double-angle formulae, we obtain
\begin{align}
f(s)   & =    -\int_0^{2\pi}   \frac{\sin(\phi)}{s}    \frac{\partial}{\partial\phi}    \left(   \frac{1}{\sqrt{ 1-2s\cos(\phi)+s^2 }}   \right) d\phi   \nonumber 
\\     
		&=    \frac{2(1+s)}{s^2}    \left[   \frac{1+s^2}{(1+s)^2} \, K\!\left(  \frac{4s}{(1+s)^2} \right)  -  E\!\left(\frac{4s}{(1+s)^2} \right)  \right], 
\label{eq:f(s)_elliptic_app}
\end{align}
where 
\begin{align*}
K(z)    =    \int_0^{\pi/2} \frac{1}{\sqrt{1-z\sin^2(\sigma)}}  \,   d \sigma 
\quad \text{and} \quad
E(z)    =    \int_0^{\pi/2} \sqrt{1-z\sin^2(\sigma)}  \,  d\sigma
\end{align*}
are the complete elliptic integrals of the first and second kind, respectively, and $\sigma=\phi/2$.

\subsection{Asymptotics of $f(s)$}\label{app:subsec:f_asymptotics}
From Eq.\ \eqref{eq:f(s)_elliptic_app}, and using the asymptotics of $K(z)$ and $E(z)$ around $s=0,1,\infty$, 
we obtain
\begin{subequations}\label{eq:f(s)_asymp_app}
\begin{align}
f(s) = \pi + \mathcal{O}(s^2)  &    \qquad \text{as} \qquad s \to 0  \label{eq:f(0)_asymp_app}, \\ 
f(s) = -2 
		\ln
		 |s-1| + \mathcal{O}(1)  & \qquad \text{as} \qquad s \to 1  \label{eq:f(1)_asymp_app}, \\ 
f(s) = \frac{\pi}{s^3} + \mathcal{O}\left(\frac{1}{s^5}\right)  & \qquad \text{as} \qquad s \to \infty  \label{eq:f(infty)_asymp_app}.  \
\end{align}
\end{subequations}

The integrand in Eq.\ \eqref{eq:f(s)_app} is undefined at $s=1$ when $\phi=0$. This behaviour is resolved by Eq.\ \eqref{eq:f(1)_asymp_app}: we see that $f(s)$ has a logarithmic singularity as $s\to 1$. This singularity is integrable in Eq.\ \eqref{eq:SLAM_Q} as long as the rest of the integrand is regular as $\omega_2 \to \omega$. This regularity holds for all cases presented in this paper.

%
%

\subsection{$f(s)$ with reciprocal argument}
\label{app:subsec:f_oneover_s}
Note that from Eqs.\ \eqref{eq:f(s)_app} and  \eqref{eq:f(s)_elliptic_app}, transforming $s\to 1/s$ gives
\begin{align}\label{eq:f_oneover_s_app}
f(1/s)  = s^3 f(s).
\end{align}

\section{Locality of power-law spectra}
\label{app:locality}

	In this appendix we carry out an analysis of the convergence of the collision integral for general power-law spectra $n_\omega = C\omega^{-x}$. We do this for completeness, and as a demonstration of the ease of analysis that the SLAM permits.

	The integral in Eq.\ \eqref{eq:Q_powerlaw} could diverge as $\omega_2 \to 0$ or $\omega_2 \to \infty$. 
	The behaviour of $f(s)$ in these ranges is noted in Appendix \ref{app:subsec:f_asymptotics}.

For $\omega_2\to \infty$, i.e.\ $s\to\infty$, we have that $f(s) \to \pi/s^3$, therefore 
$Q \propto \int^{\infty}  \omega_2^{1-3/2-2x}(1-\omega_2^{x-1})\ d \omega_2$,
which is convergent for $x>\max\{1/4,-1/2\}=1/4$. 

For $\omega_2\to 0$, i.e.\ $s\to 0$, we have $f(s) \to \pi$, and so
$Q \propto \int_0  \omega_2^{1-2x}(1-\omega_2^{x-1})\ d \omega_2$,
which is convergent for $x \leq1$. 

For either choice of $\omega_2$, the waveaction and energy equipartition spectra lead to convergence of the collision integral, since the factor $(\partial_{\omega} n_{\omega}^{-1} - \partial_{\omega_2} n_{\omega_2}^{-1})$ in Eq.\ \eqref{eq:SLAM_Q} vanishes exactly for any RJ spectrum \eqref{eq:RJ}. 

Thus, for $\omega_2\to \infty$, the SLAM converges for power-law spectra with spectral index $x \in \{(1/4, \infty) \cup 0\}$, and for $\omega_2\to 0$ it converges for $x \in (-\infty,1]$. Otherwise, the SLAM is divergent. These convergence (green) and divergence (red) zones are indicated in Fig.\ \ref{fig:indices_fluxes_convergence}(b), for the two choices of $\omega_2 \to \infty$ or $0$, above and below the $x$ axis respectively. The thin green (convergence) strips around the thermodynamic spectra are indicative only, and in reality shrink to the single points $x=0,1$.

We see that the KZ energy cascade spectrum, with $x=1/2$, gives convergence of the SLAM, whereas the KZ waveaction cascade spectrum, with $x=1/6$, gives divergence as $\omega_2\to\infty$
(These results were found in Sec.\ \ref{subsec:KZspectra}, but now we see them in their full context of convergence or divergence on general power-law spectra).
	
	Note that analysing the locality of general power-law spectra is made possible in the SLAM precisely because of the \emph{semilocality} manifested by the interaction coefficient of the SHE. This analysis is not possible when working with a DAM, because in order to construct a DAM one assumes (without proof) from the outset that interacting waves are superlocal.

\bibliography{bib_SHE_SLAM_2D}

\end{document}